\documentclass[a4paper,twocolumn,11pt,unpublished]{quantumarticle}
\pdfoutput=1
\usepackage[utf8]{inputenc}
\usepackage[english]{babel}
\usepackage[T1]{fontenc}
\usepackage[numbers,sort&compress]{natbib}
\usepackage{amsmath,amssymb,amsfonts}
\usepackage{hyperref}
\usepackage{multirow}

\usepackage{tikz}
\usepackage{lipsum}

\newcommand{\ket}[1]{|#1\rangle}
\newcommand{\bket}[1]{|\boldsymbol{#1}\rangle}
\newcommand{\bfket}[1]{|\mathbf{#1}\rangle}

\newcommand{\bfbra}[1]{\langle \mathbf{#1} |}
\newcommand{\al}{\alpha_{\ell}}
\newcommand{\ald}{\tilde{\alpha}_{\ell}}
\newcommand{\lal}{\lambda_{\ell}}
\newcommand{\lald}{\tilde{\lambda}_{\ell}}
\newcommand{\Sl}{\Sigma_{\ell}}
\newcommand{\nuamp}{\nu_{\mathrm{amp}}}
\newcommand{\creg}{\mathsf{C}}
\newcommand{\dreg}{\mathsf{D}}
\newcommand{\ereg}{\mathsf{E}}
\newcommand{\anc}{\mathsf{a}}
\newcommand{\ganc}{\mathsf{g}}

\begin{document}

\title{Quantum state preparation with multiplicative amplitude transduction}

\author{Yutaro Iiyama}
\affiliation{International Center for Elementary Particle Physics, The University of Tokyo, Tokyo, Japan}
\email{iiyama@icepp.s.u-tokyo.ac.jp}
\orcid{0000-0002-8297-5930}
\maketitle

\begin{abstract}
Quantum state preparation is an important class of quantum algorithms that is employed as a black-box subroutine in many algorithms, or used by itself to generate arbitrary probability distributions. We present a novel state preparation method that utilizes less quantum computing resource than the existing methods. Two variants of the algorithm with different emphases are introduced. One variant uses fewer qubits and no controlled gates, while the other variant potentially requires fewer gates overall. A general analysis is given to estimate the number of qubits necessary to achieve a desired precision in the amplitudes of the computational basis states. The validity of the algorithm is demonstrated using a prototypical problem of generating Ising model spin configurations according to its Boltzmann distribution.
\end{abstract}

\section{Introduction}

Quantum state preparation is a digital quantum computing algorithm that transforms the state of the quantum computer from a trivially realizable initial state into an arbitrary superposition of computational basis states. The list of quantum algorithms that depend on some form of state preparation as a subroutine includes, but is not limited to, linear system solvers~\cite{Harrow_2009,doi:10.1137/16M1087072}, principal component analysis~\cite{pca}, and discrete-time quantum walk~\cite{qwalk}. Engineering of quantum state is also interesting in itself, since a state preparation subroutine immediately followed by measurements of the qubit states in the computational basis acts as a random number generator that samples from a distribution specified by the modulus squared of the amplitude of the basis states~\cite{Zoufal2019}.

The first concrete state preparation algorithm was presented in Ref. \cite{PhysRevLett.85.1334}. In this method, desired amplitudes of the computational basis states are encoded into the phase of $Y$ rotation of an ancilla qubit, which is then projected onto $Z$ eigenstates. Since only one of the projections are accompanied by the correct amplitude, a generalization of quantum search algorithm~\cite{PhysRevLett.79.325} is employed to amplify the desired projection and ``purify'' the overall state.

Reference \cite{PhysRevLett.122.020502} introduced a method to synthesize the amplitude without phase rotation. Because the method in Ref. \cite{PhysRevLett.85.1334} would in general require a quantum computation of the arcsine function, which is a highly complex operation by itself~\cite{hner2018optimizing}, elimination of phase rotation represents a major simplification of the state preparation algorithm. Amplitude transduction in Ref. \cite{PhysRevLett.122.020502} also results in a superposition of desired and undesired states, which can then be plugged into the amplitude amplification routine similar to Ref. \cite{PhysRevLett.85.1334}. 

Recently, a drastically different approach to state preparation was introduced in Ref. \cite{Zoufal2019}, whereby state amplitudes are approximated parametrically by applying alternating sets of rotation and entangling gates onto a fixed initial state. The angles in the rotation gates are free parameters to be machine-learned from existing examples by using this parametric quantum circuit as the generator in a generative adversarial network setup.

In this work, we present a novel amplitude transduction method that is both simpler and uses less qubits than Ref. \cite{PhysRevLett.122.020502}. The key point of the method is to record the logarithm of the target amplitude in the data register, from which the amplitude can be synthesized multiplicatively.

This paper is organized as follows. In Section~\ref{sec:existing}, we briefly review the existing state preparation algorithms in Refs. \cite{PhysRevLett.85.1334} and \cite{PhysRevLett.122.020502}. Section~\ref{sec:multiplicative} introduces the multiplicative amplitude transduction algorithm and discusses its properties. In Section~\ref{sec:ising}, we demonstrate the algorithm in a concrete use case of a simple Ising model computation. Finally, a summary and future prospects are given in Section~\ref{sec:conclusions}.

\section{Existing methods}
\label{sec:existing}

Let us first introduce the setup of the problem of state preparation. Suppose a quantum register of $n$ qubits, which we call the configuration register $\creg$. The Hilbert space of the $i$th qubit is spanned by the two eigenstates of operator $Z$ denoted as $\ket{0}_i$ and $\ket{1}_i$. Let $N = 2^n$. Computational basis states $\{\ket{\ell}_{\creg}\, |\, \ell=0,\dots,N-1\}$ of the register are defined as
\begin{equation} \label{eqn:comp_basis}
    \ket{\ell}_{\creg} := \bigotimes_{i=0}^{n-1} \ket{\ell^{(i)}}_{i}, \thickspace \text{where} \thickspace \ell = \sum_{i=0}^{n-1} \ell^{(i)}2^i.
\end{equation}
Fix the initial state of the register to $\bfket{i}_{\creg} := \ket{0}_{\creg}$. Given an array $\{\al \in [0,1] \,|\, \ell=0,\dots,N-1\}$\footnote{Note that the constraint on $\al$ does not represent a loss of generality, as noted in Ref. \cite{PhysRevLett.122.020502}, because general complex amplitudes can be represented in polar form. We let $\al$ represent the modulus of the amplitudes, while their phases can be directly transduced from controlled phase operations. The modulus is obviously finite and can therefore be normalized by its maximum, or if that is unknown, some upper bound value.}, the goal of a state preparation routine is to transform the initial state into the target state
\begin{equation}
    \bfket{f}_{\creg} := \frac{1}{\mathcal{A}} \sum_{\ell=0}^{N-1} \al \ket{\ell}_{\creg}, \thickspace \text{where} \thickspace \mathcal{A} = \sqrt{\sum_{\ell=0}^{N-1}\al^2}.
\end{equation}

In the amplitude synthesis method of Ref. \cite{PhysRevLett.85.1334}, the configuration register is accompanied by a single ancilla qubit $\anc$ initialized to $\ket{0}_{\anc}$. The method then assumes a unitary operator $\mathfrak{R}$ whose action on $\creg \otimes \anc$ is
\begin{equation}
    \mathfrak{R}\ket{\ell}_{\creg} \ket{0}_{\anc} = \ket{\ell}_{\creg} \left(\cos\theta_{\ell} \ket{0}_{\anc} + \sin\theta_{\ell} \ket{1}_{\anc} \right),
\end{equation}
where $\theta_{\ell} = \arccos \al$. Exact implementation of $\mathfrak{R}$ is problem-specific, but the author posits that it can be generally realized through conditional phase rotation circuits. Let $H_{\mathsf{X}}$ represent Hadamard gates being applied to all qubits of register $\mathsf{X}$. Then,
\begin{equation} \label{eqn:grover}
    \begin{split}
    \mathfrak{R} H_{\creg} \ket{0}_{\creg} \ket{0}_{\anc} & = \frac{1}{\sqrt{N}}\sum_{\ell=0}^{N-1}\mathfrak{R}\ket{\ell}_{\creg} \ket{0}_{\anc} \\
    & = \frac{\mathcal{A}}{\sqrt{N}} \bfket{f}_{\creg} \ket{0}_{\anc} + \Omega \bket{\omega}_{\creg} \ket{1}_{\anc},
    \end{split}
\end{equation}
i.e., one can synthesize a state vector proportional to $\bfket{f}_{\creg}$ along the projection onto $\ket{0}_{\anc}$. Here and throughout the remainder of this paper, we denote uninteresting ``byproduct'' states and their norms by $\bket{\omega}$ and $\Omega$, respectively. The exact value of $\mathcal{A}/\sqrt{N} = \sqrt{1 - \Omega ^2} < 1$ in Eq.~\eqref{eqn:grover} depends on the given $\{\al\}$, but is in general expected to be $\sim 1/\sqrt{2}$. However, one can amplify the amplitude of the target state using the method described later in this section.

It must be noted that this seemingly simple algorithm generally consumes substantial resource, in terms of the number of both qubits and gates. That is, unless the functional form of $\{\al\}$ is such that a simple quantum circuit to directly compute $\theta_{\ell}$, instead of $\al$, from $\ell$ is known, $\mathfrak{R}$ entails a general quantum circuit for calculation of the arcsine function, which requires $\mathcal{O}(100)$ qubits and $\mathcal{O}(10^4)$ Toffoli gates~\cite{hner2018optimizing}.

Method of Ref. \cite{PhysRevLett.122.020502} aims to reduce the complexity by replacing the $\theta_{\ell}$-rotation by comparisons and basis counting. This method uses two additional $d$-qubit registers, $\dreg$ and $\ereg$, and an ancilla bit $\ganc$. For each of $\dreg$ and $\ereg$, $2^d$ computational basis states are given by definitions similar to Eq.~\eqref{eqn:comp_basis}. A unitary operator $\mathfrak{A}$ is assumed to exist with action
\begin{equation}
    \mathfrak{A} \ket{\ell}_{\creg} \ket{0}_{\dreg} = \ket{\ell}_{\creg} \ket{\ald}_{\dreg},
\end{equation}
where $\ald = \left\lfloor 2^d \al \right\rfloor$. The register $\ereg$ is prepared in a uniform superposition $H_{\ereg} \ket{0}_{\ereg} = 1/\sqrt{2^d} \sum_{x} \ket{x}_{\ereg}$. A comparison operator $\mathfrak{C}$ acts on $\dreg \otimes \ereg \otimes \ganc$ by setting the value of $\ganc$ to 1 for bases of $\ereg$ with index greater than or equal to the value in $\dreg$:
\begin{multline} \label{eqn:count}
    \mathfrak{C} \ket{\ald}_{\dreg} \frac{1}{\sqrt{2^d}}\sum_{x=0}^{2^d-1} \ket{x}_{\ereg} \ket{0}_{\ganc} = \\
    \ket{\ald}_{\dreg} \frac{1}{\sqrt{2^d}} \left[ \sum_{x=0}^{\ald - 1} \ket{x}_{\ereg} \ket{0}_{\ganc} + \sum_{x=\ald}^{2^d-1} \ket{x}_{\ereg} \ket{1}_{\ganc} \right].
\end{multline}
Then, another application of $H_{\ereg}$ to Eq.~\eqref{eqn:count} results in $\ald$ instances of term $1/\sqrt{2^d} \ket{0}_{\ereg} \ket{0}_{\ganc}$ from the first sum in the bracket. Therefore,
\begin{equation}
    \begin{split}
    &H_{\ereg}\mathfrak{C}\mathfrak{A}H_{\ereg} H_{\creg} \ket{0}_{\creg} \ket{0}_{\dreg} \ket{0}_{\ereg} \ket{0}_{\ganc} \\
    & = \frac{1}{\sqrt{N}} \sum_{\ell=0}^{N-1} \frac{\ald}{2^d} \ket{\ell}_{\creg} \ket{\ald}_{\dreg} \ket{0}_{\ereg} \ket{0}_{\ganc} + \bket{\omega} \\
    & =: \frac{\tilde{\mathcal{A}}}{\sqrt{N}} \ket{\tilde{\mathbf{f}}}_{\creg}  \ket{\ald}_{\dreg} \ket{0}_{\ereg} \ket{0}_{\ganc} + \bket{\omega},
    \end{split}
\end{equation}
where $\tilde{\mathcal{A}} = \sqrt{\sum_{\ell} (\ald/2^d)^2}$. State 
\begin{equation}
    \ket{\tilde{\mathbf{f}}}_{\creg} = \frac{1}{\tilde{\mathcal{A}}} \sum_{\ell=0}^{N-1} \frac{\ald}{2^d} \ket{\ell}_{\creg}
\end{equation}
 is identical to $\bfket{f}_{\creg}$ up to $\mathcal{O}(2^{-d})$. Because $\bket{\omega}$ does not contain terms with $\ket{0}_{\ereg} \ket{0}_{\ganc}$, the target state is synthesized along the projection onto $\ket{0}_{\ereg} \ket{0}_{\ganc}$, with its pre-amplification amplitude within $\mathcal{O}(2^{-d})$ to that of the first method.

According to Ref. \cite{PhysRevLett.122.020502}, the comparison operation $\mathfrak{C}$ requires $d$ Toffoli gates and additional $d$ qubits. Therefore, this amplitude synthesis method is considerably more realistic to be deployed on a near-term quantum device than the first one.

In both amplitude synthesis methods, one is left with a superposition of the desired final state and undesired byproduct states. An important contribution of Ref. \cite{PhysRevLett.85.1334} was to show the algorithm for amplifying the magnitude of the amplitude of the desired state in such a superposition, regardless of the amplitude synthesis method.

The outline of the amplification algorithm is as follows. First, consider the entire amplitude synthesis procedure as an application of a single unitary $U$ to the ``source'' state $\bfket{i} \bfket{s}$, where $\bfket{s}$ represents the initial state of the ancillary registers. The synthesized term to amplify is then $\sum_{\ell} U_{\ell,\mathbf{t};\mathbf{i},\mathbf{s}} \ket{\ell} \bfket{t}$, where the ket $\bfket{t}$ defines the subspace in which the target state lies. The matrix element $U_{\ell,\mathbf{t};\mathbf{i},\mathbf{s}}$ is proportional to $\al$. Crucially, $\bfket{s}$ and $\bfket{t}$ must be ``simple'' states such that selective phase inversion operations $I_{\mathbf{s}} = I - 2\bfket{s}\bfbra{s}$ and $I_{\mathbf{t}} = I - 2\bfket{t}\bfbra{t}$, with $I$ being the identity operator, can be implemented using only a few gates. If this condition is satisfied, which is the case for the two amplitude synthesis methods above, then it was proven in Ref. \cite{PhysRevLett.85.1334} that the combined operation
\begin{equation} \label{eqn:Q}
    Q = -I_{\mathbf{s}} U^{-1} I_{\mathbf{t}} U
\end{equation}
rotates the state vector within a two-dimensional subspace spanned by $\bfket{i} \bfket{s}$ and $\sum_{\ell} U_{\ell,\mathbf{t};\mathbf{i},\mathbf{s}} U^{-1}\ket{\ell}\bfket{t}$. Using the pre-amplification norm of the target state $u = \sqrt{\sum_{\ell} |U_{\ell,\mathbf{t};\mathbf{i},\mathbf{s}}|^2}$, the angle of rotation is approximately $4u/\pi$. Therefore, for an integer $\nuamp \sim \pi / (4u)$,
\begin{equation} \label{eqn:amplification}
    U Q^{\nuamp} \bfket{i} \bfket{s} = \frac{1}{\mathcal{A}'} \sum_{\ell=0}^{N-1} \al \ket{\ell} \bfket{t} + \epsilon \bket{\omega},
\end{equation}
where $\mathcal{A}' \sim \mathcal{A}$ and $|\epsilon| \ll 1$.

\section{Multiplicative amplitude transduction}
\label{sec:multiplicative}

We now describe our approach to amplitude synthesis. The method is similar to that in Ref. \cite{PhysRevLett.122.020502} and proceeds by recording the value of a function of $\ell$ into a $d$-qubit register $\dreg$, from which the amplitude is transduced. The difference from the previous method is that the function is not $\al$ itself but rather its logarithm, and the amplitude is computed multiplicatively. Two variants of the algorithm will be shown. One variant requires no controlled gates in the amplitude transduction process but has a smaller pre-amplification norm of the target state, and the other achieves pre-amplification norm that is numerically equivalent to the methods in the previous section.

\subsection{Algorithm}

Let
\begin{equation} \label{eqn:lal}
    \lal := -\log_{\gamma} \al
\end{equation}
for an arbitrary $\gamma > 1$. For simplicity, assume for now that we are given $\al > 0$ for all $\ell$. The size of the $\dreg$ register $d$ must be such that $\max_{\ell} \lal < 2^d$. Then define
\begin{equation}
    \lald := \left\lfloor \lal \right\rfloor,
\end{equation}
and assume there is a unitary operator $\mathfrak{L}$ with action
\begin{equation} \label{eqn:v1}
    \mathfrak{L} \ket{\ell}_{\creg} \ket{0}_{\dreg} = \ket{\ell}_{\creg} \ket{\lald}_{\dreg}.
\end{equation}
For later convenience, define the binary expansion of $\lald$ by
\begin{equation}
    \lald = \sum_{k=0}^{d-1} \lald^{(k)}2^k
\end{equation}

In the first variant, we apply to $\dreg$ an operator $\mathfrak{T}_1$, which rotates each qubit of $\dreg$ by the standard single-qubit rotation gate $R_{y}(-2\phi_k)$. The angle $\phi_k$ for the $k$th qubit $(k=0, \dots, d-1)$ is given by
\begin{equation} \label{eqn:phik}
    \phi_k := \arctan \left(\gamma^{-2^k}\right).
\end{equation}
Then,
\begin{equation} \label{eqn:t1}
    \begin{split}
        & \mathfrak{T}_1 \ket{\lald}_{\dreg} \\
        = & \bigotimes_{k=0}^{d-1} \cos \phi_k \left[ (\tan \phi_k)^{\lald^{(k)}} \ket{0}_k + (-\tan \phi_k)^{1 - \lald^{(k)}} \ket{1}_k \right] \\
    = & \left( \prod_{k=0}^{d-1} \cos \phi_k \right) \left[ \gamma^{-\lald} \ket{0}_{\dreg} + \Omega_{\dreg} \bket{\omega}_{\dreg} \right].
    \end{split}
\end{equation}
Denoting the constant $\prod_{k} \cos \phi_k$ as $\Phi$, Eq.~\eqref{eqn:t1} leads to
\begin{equation}
\begin{split}
\mathfrak{T}_1\mathfrak{L}H_{\creg} \ket{0}_{\creg} \ket{0}_{\dreg} & = \Phi \frac{1}{\sqrt{N}} \sum_{\ell=0}^{N-1} \gamma^{-\lald} \ket{\ell}_{\creg} \ket{0}_{\dreg} + \Omega \bket{\omega} \\
 &=: \Phi \frac{\bar{\mathcal{A}}}{\sqrt{N}} \ket{\bar{\mathbf{f}}}_{\creg} \ket{0}_{\dreg} + \Omega \bket{\omega},
\end{split}
\end{equation}
where $\tilde{\mathcal{A}} = \sqrt{\sum_{\ell} \gamma^{-2\lald}}$. State 
\begin{equation}
    \ket{\bar{\mathbf{f}}}_{\creg} = \frac{1}{\bar{\mathcal{A}}} \sum_{\ell=0}^{N-1} \gamma^{-\lald} \ket{\ell}_{\creg}
\end{equation}
is an approximation of $\bfket{f}_{\creg}$ up to $\mathcal{O}(2^{-d} \ln \gamma)$. We have thus transduced the desired amplitudes using only single-qubit gates in the $\dreg$ register. However, the constant factor $\Phi$ is asymptotically $\sim \sqrt{\gamma - 1}$ for large $d$, which for $\ln \gamma \ll 1$ implies that a large number of amplitude amplification iterations is necessary in Eq.~\eqref{eqn:amplification}. We refer to this version of the algorithm as the \textit{direct} variant hereafter.

In the second variant, we invoke a second $d$-qubit register $\ereg$ in the initial state $\ket{0}_{\ereg}$, and apply to $\dreg \otimes \ereg$ an operator $\mathfrak{T}_2$, which consists of rotation $R_{y}(2\psi_k)$ on each qubit of $\ereg$ controlled by the corresponding qubit in $\dreg$. The angle $\psi_k$ for the $k$th qubit is given by
\begin{equation} \label{eqn:psik}
    \psi_k := \arccos \left(\gamma^{-2^k}\right).
\end{equation}
Then,
\begin{equation} \label{eqn:t2}
    \begin{split}
        & \mathfrak{T}_2 \ket{\lald}_{\dreg} \ket{0}_{\ereg} \\
        = & \bigotimes_{k=0}^{d-1} \ket{\lald^{(k)}}_{\dreg} \left[ (\cos \psi_k)^{\lald^{(k)}} \ket{0}_k + \sin (\lald^{(k)} \psi_k) \ket{1}_k \right] \\
    = & \gamma^{-\lald} \ket{\lald}_{\dreg} \ket{0}_{\ereg} + \Omega_{\dreg \otimes \ereg} \bket{\omega}_{\dreg \otimes \ereg}.
    \end{split}
\end{equation}
Therefore,
\begin{equation}
    \mathfrak{T}_2\mathfrak{L}H_{\creg} \ket{0}_{\creg} \ket{0}_{\dreg} \ket{0}_{\ereg} = \frac{\bar{\mathcal{A}}}{\sqrt{N}} \ket{\bar{\mathbf{f}}}_{\creg} \ket{\lald}_{\dreg} \ket{0}_{\ereg} + \Omega \bket{\omega}.
\end{equation}
The byproduct state $\bket{\omega}$ does not contain $\ket{0}_{\ereg}$, which defines the projection the target state lies in. In this version of the algorithm, which is referred to as the \textit{controlled} variant, the pre-amplification norm of the target state is numerically equivalent to those of the methods in the previous section.

\subsection{Observations}

An immediate question that arises regarding this method is the feasibility of implementing the operation $\mathfrak{L}$ as a quantum circuit. The answer is obviously problem-specific, but we claim that in the most general case where $\ald$ has to be approximated by a series expansion in $\ell$, expanding $\lald$ instead does not incur significant additional complexity to the circuit. In other words, $\mathfrak{L}$ should be in general feasible if $\mathfrak{A}$ is. There are nevertheless certain forms of $\al$ where expression using $\ell$ is significantly simpler for either $\ald$ or $\lald$; one should examine the problem at hand and decide the best approach for amplitude transduction. Later, we present the latter case as a demonstration of our algorithm.

One of the most prominent features of this algorithm is its low quantum computational cost and, in the direct variant, fault tolerance in the amplitude transduction step. Note that the rotation angles in Eqs.~\eqref{eqn:phik} and \eqref{eqn:psik} are determined solely by the value of $\gamma$, which is a priori chosen by the implementer of the algorithm. Therefore, the only quantum gates involved in the transduction step are, in the direct variant, the $d$ instances of separate single-qubit rotations by constant angles, and in the controlled variant, corresponding single-controlled rotations. There is also no ``hidden'' qubit requirement like the additional $d$ qubits needed to perform the comparison in Ref. \cite{PhysRevLett.122.020502}. It is also worth mentioning that rotations about any axis in the $X$-$Y$ plane are equally usable for both variants.

The size $d$ of the $\dreg$ and $\ereg$ registers is determined by the desired precision of approximation of $\bfket{f}_{\creg}$ by $\ket{\bar{\mathbf{f}}}_{\creg}$. First, note that $\al \ll 1$ implies that state $\ket{\ell}_{\creg}$ contributes negligibly to $\bfket{f}_{\creg}$, rendering the exact value of $\al$ unimportant. One can thus consider implementing $\mathfrak{L}$ so that for $\al$ below some cutoff $\epsilon$, the value of $\lald$ is fixed at $2^{d} - 1$. Second, in general, choosing $\gamma = e^{\delta}$ allows an approximation of any $\al$ within a relative precision of $\delta$. Combining the two relations,
\begin{equation} \label{eqn:precision}
    2^d > -\log_{\gamma} \epsilon = \frac{-\ln \epsilon}{\delta}.
\end{equation}
Therefore, for example, for $\delta = \epsilon = 0.001$, $d = 13$ qubits are needed. This is the exact number of qubits necessary for amplitude transduction in the direct variant of our algorithm, while the controlled variant requires $2d = 26$ qubits. In contrast, in the algorithm in Ref. \cite{PhysRevLett.122.020502}, where the cutoff and absolute precision of $\al$ are equivalent, a cutoff of 0.001 requires $d = 10$ qubits per register for three registers.

So far we have assumed that $\al > 0$. By letting $\lald$ for $\al < \epsilon$ saturate as above, we are in principle able to also handle the case $\al = 0$. If we need to ensure that state $\ket{\ell}_{\creg}$ with genuinely null $\al$ does not appear in $\ket{\bar{\mathbf{f}}}_{\creg}$, one possible workaround is to write the NAND of all qubits in $\dreg$ to an ancilla qubit. For non-saturated states, $R_{y}$ rotations can be controlled on this qubit, while for the saturated states, NOT gates anti-controlled on this qubit can be used instead of the rotations, setting the coefficient of $\ket{0}_{\dreg}$ or $\ket{0}_{\ereg}$ to 0. As a collateral effect of this procedure, all $\ket{\ell}_{\creg}$ with $\al < \epsilon$ will vanish from the final state, which should not be a problem if approximation by the cutoff prescription is acceptable in the first place.

\section{Demonstration with an Ising model sampler}
\label{sec:ising}

As a demonstration of the algorithm introduced in the previous section, we consider a simple quantum circuit that generates random numbers according to a probability distribution specified through the amplitudes of the computational basis states of the $\creg$ register. The circuit comprises one amplitude synthesis block and zero or more amplification blocks, where each of the latter actually contains an amplitude synthesis block and its inverse, as seen in Eq.~\eqref{eqn:Q}. The circuit is then terminated with measurements of the $\creg$ register and $\dreg$ or $\ereg$ register, depending on the variant of the algorithm to be demonstrated. A sequence of desired random numbers are obtained in the readout of $\creg$, when the circuit is executed multiple times and results are kept only when the readout of $\dreg$ or $\ereg$ is 0.

We chose the Boltzmann distribution of the Ising model as the probability distribution to sample from. We let each qubit in the configuration register $\creg$ correspond to a spin site in the Ising model, with values 0 and 1 of the qubit corresponding to down and up states of the spin. The quantum circuit aims to synthesize the state
\begin{equation}
    \bfket{f}_{\creg} = \frac{1}{\sqrt{\mathcal{Z}}} \sum_{\ell = 0}^{N-1} \exp (-\beta E_{\ell}/2) \ket{\ell}_{\creg},
\end{equation}
where $\beta$ is the inverse temperature, $E_{\ell}$ is the energy of the spin configuration $\ell$, and $\mathcal{Z}$ is the partition function of the system. Clearly, this is a problem particularly suited for our algorithm, because of the explicit exponentiation in the amplitudes. The configuration $\ell$ should be generated with frequency proportional to $\exp (-\beta E_{\ell})$.

To maximally simplify the problem, let us consider a two-dimensional isotropic square lattice Ising model with a periodic boundary condition and no external field, represented by the Hamiltonian
\begin{equation}
    H = -J \sum_{\langle i, j \rangle} s_i s_j.
\end{equation}
In the expression, the summation is over all $2N$ nearest-neighbor spin pairs, and $s_i = \pm 1$ is the spin value at site $i$. Denoting the number of neighboring spin pairs with opposing spins in the configuration $\ell$ by $\Sl$,
\begin{equation}
    \sum_{\langle i, j \rangle} s_i s_j = 2 N - 2 \Sl,
\end{equation}
and therefore
\begin{equation}
    \exp (-\beta E_{\ell}/2) = \exp \left(\beta J (N - \Sl) \right).
\end{equation}
Factoring out the constant $\exp (\beta J N)$, we define $\al = \exp (-\beta J \Sl)$, which satisfies the constraint $\al \in [0, 1]$. Since $\Sl$ is an even number for all $\ell$, we can set $\gamma = \exp (-2\beta J)$ in Eq.~\eqref{eqn:lal} to obtain $\lal = \Sl/2$. In this setup, $\lal$ is an integer, implying $\lald = \lal$, and therefore the precision analysis in Eq.~\eqref{eqn:precision} does not apply. The size of $\dreg$ is determined by $2^d > \max_{\ell}\Sl/2$.

\begin{figure}
    \centering
    \includegraphics[width=0.48\textwidth]{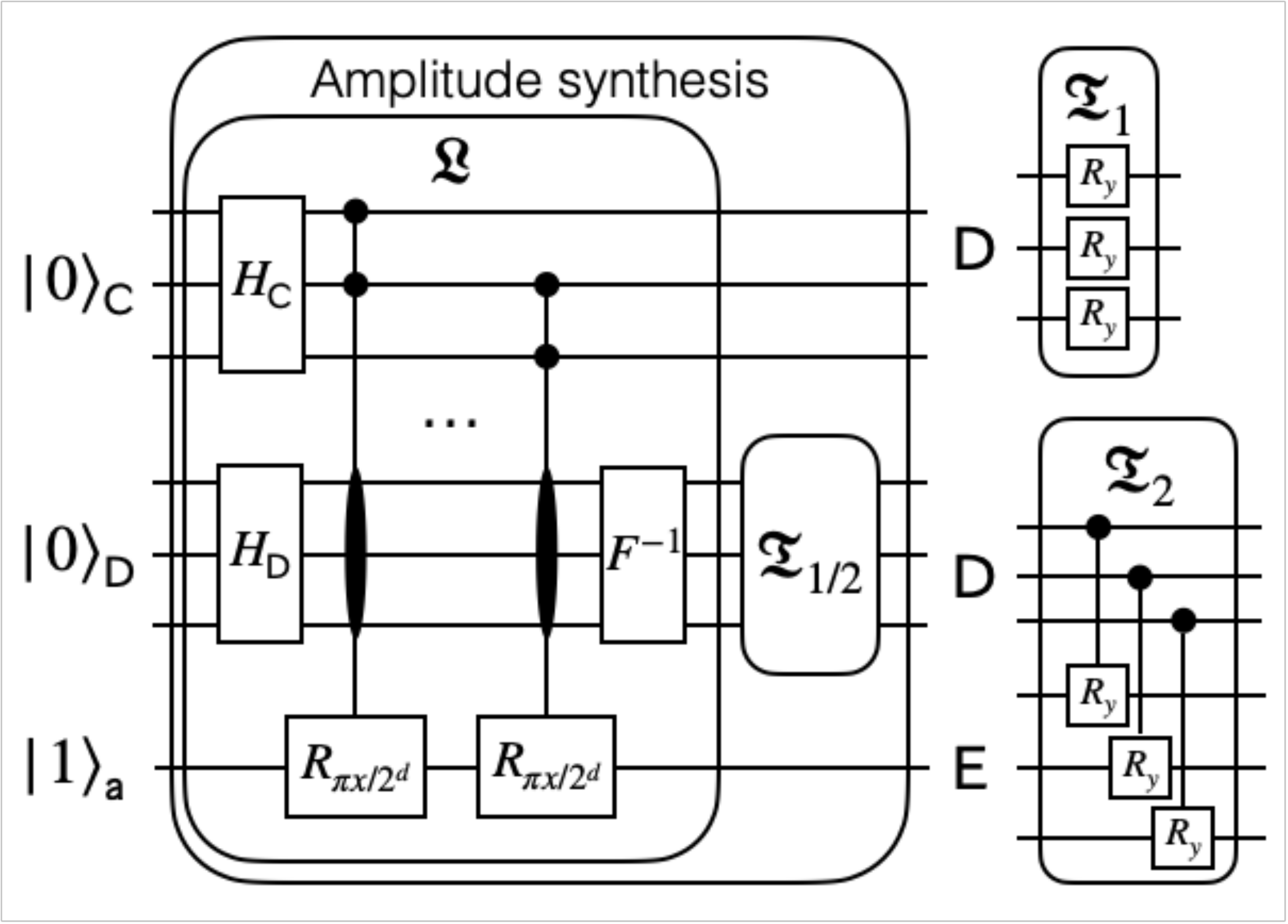}
    \caption{Schematic depiction of the amplitude synthesis circuit. In the inner left block corresponding to the $\mathfrak{L}$ operator, each opposing spin pair in the $\creg$ register contribute a phase shift of $\pi x/2^d$ to the basis $x$ of $\dreg$. The accumulated phase is converted to half the number of opposing spin pairs through inverse Fourier transform. The amplitudes of the spin configurations are transduced via the $\mathfrak{T}_1$ or $\mathfrak{T}_2$ operator, drawn separately on the right.}
    \label{fig:pair_counting}
\end{figure}

The operator $\mathfrak{L}$ is implemented using a similar logic to the well-known quantum phase estimation algorithm (Fig.~\ref{fig:pair_counting}). First, $\dreg$ is prepared in a uniform superposition and an ancilla qubit $\anc$ is introduced in state $\ket{1}_{\anc}$. Then, using a multiple-control phase rotation gate ($R_{\phi}$), the phase of $\anc$ is shifted by $\pi x / 2^d$ for XOR of every pair of qubits in $\creg$ representing neighboring spins, for every $x$ in the superposition in the $\dreg$ register. After all pairs are processed, the phase of $\ket{\ell}_{\creg} \ket{x}_{\dreg}$ is $\exp (\pi i x \Sl / 2^d)$. Finally, an inverse Fourier transform $F^{-1}$ is applied on $\dreg$ to obtain
\begin{multline}
    \ket{\ell}_{\creg} F^{-1} \left[ \frac{1}{2^{d-1}} \sum_{x=0}^{2^d -1} \exp \left(\frac{\pi i x \Sl}{2^d} \right)\ket{x}_{\dreg} \right] = \\
    \ket{\ell}_{\creg} \ket{\Sl/2}_{\dreg}.
\end{multline}

The circuits for $2 \times 2$, $3 \times 3$, and $4 \times 4$ spin lattices with $\beta J=0.1$ are implemented in Python using the Qiskit library, for both variants of the amplitude transduction algorithm, and executed on the built-in QASM simulator. Each circuit is executed for $2^{17}$ times (``shots''). Table~\ref{tab:simulations} summarizes the circuit parameters, number of amplification iterations $\nuamp$, squares of pre- and post-amplification norms of the target state ($u^2$ and $\mathcal{A}'^2$), and the efficiency $\mathcal{E}$ of observing 0 in $\dreg$ or $\ereg$ of the simulation experiments. This efficiency should be identical to $\mathcal{A}'^2$ within statistical uncertainty. The gate counts in the table are for a single pass of the circuit corresponding to $\mathfrak{T}_1$ or $\mathfrak{T}_2$, composed of CNOT, frame change, and $X_{\pi / 2}$ gates, assuming the topology of the IBM Quantum Experience hardware \texttt{ibmq\_cambridge}. These gate counts are in principle unspecific to the experiment and depend solely on $d$. However, in practice, the specific combination of the circuit structure and the topology of the underlying hardware may result in circuits with identical $d$ values having different gate counts.

\begin{table}
    \centering
    \scriptsize
    \setlength{\tabcolsep}{3pt}
    \caption{Summary of circuit simulations for two variants of the amplitude transduction algorithm for three Ising lattice sizes. Gate counts are for one execution of the transduction algorithm, assuming the topology and the available gates of \texttt{ibmq\_cambridge}, without optimization. See Sections \ref{sec:existing} and \ref{sec:ising} for symbol definitions.}
    \label{tab:simulations}
    \begin{tabular}{c|cccccccc}
    \multirow{2}{*}{Model} & \multicolumn{2}{c}{Qubits} & \multicolumn{2}{c}{Gates} & \multirow{2}{*}{$\nuamp$} & \multirow{2}{*}{$u^2$} & \multirow{2}{*}{$\mathcal{A}'^2$} & \multirow{2}{*}{$\mathcal{E}$} \\
    & Total & $d$ & CNOT & $X_{\pi/2}$ & & \\
    \hline
    \multicolumn{9}{c}{Direct variant simulations} \\
    $2 \times 2$ & 8 & 3 & 0 & 8 & 2 & 0.167 & 0.738 & 0.743 \\
    $3 \times 3$ & 13 & 3 & 0 & 8 & 3 & 0.063 & 0.960 & 0.961 \\
    $4 \times 4$ & 22 & 5 & 0 & 12 & 6 & 0.016 & 0.996 & 0.995 \\
    \multicolumn{9}{c}{Controlled variant simulations} \\
    $2 \times 2$ & 11 & 3 & 33 & 14 & 1 & 0.487 & 0.539 & 0.535 \\
    $3 \times 3$ & 16 & 3 & 24 & 14 & 2 & 0.182 & 0.650 & 0.650 \\
    $4 \times 4$ & 27 & 5 & 43 & 22 & 4 & 0.048 & 0.837 & 0.837
    \end{tabular}
\end{table}

For each lattice size, the controlled variant synthesizes a state with a greater value of $u$ compared to the direct variant, resulting in fewer amplification iterations. However, in these specific examples, the controlled variant has $u \sim \mathcal{O}(1)$, and the granularity of the amplification too large, resulting in the direct variant with finer amplification steps achieving $\mathcal{A}'$ closer to unity.

The $2 \times 2$ circuits for both variants of the amplitude transduction algorithm are also run on \texttt{ibmq\_cambridge}. To limit the circuit depth, the circuits contain only a single routine of amplitude transduction, i.e., no amplitude amplification is performed. Table~\ref{tab:experiments} summarizes the circuit depth and $\mathcal{E}$ in the $\dreg$ or $\ereg$ register in $2^{13}$ shots for the two variants.

\begin{table}
    \centering
    \caption{Summary of experiments on \texttt{ibmq\_cambridge} using the circuits for the $2 \times 2$ Ising lattice with direct and controlled variants of the amplitude transduction algorithm. No amplitude amplification is performed. The values of $u^2$ are repeated from Table~\ref{tab:simulations}.}
    \label{tab:experiments}
    \begin{tabular}{c|ccc}
    Variant & Depth & $u^2$ & $\mathcal{E}$ \\
    \hline
    Direct & 578 & 0.167 & 0.16 \\
    Controlled & 582 & 0.487 & 0.27
    \end{tabular}
\end{table}

\begin{figure}[t]
    \centering
    \includegraphics[width=0.48\textwidth]{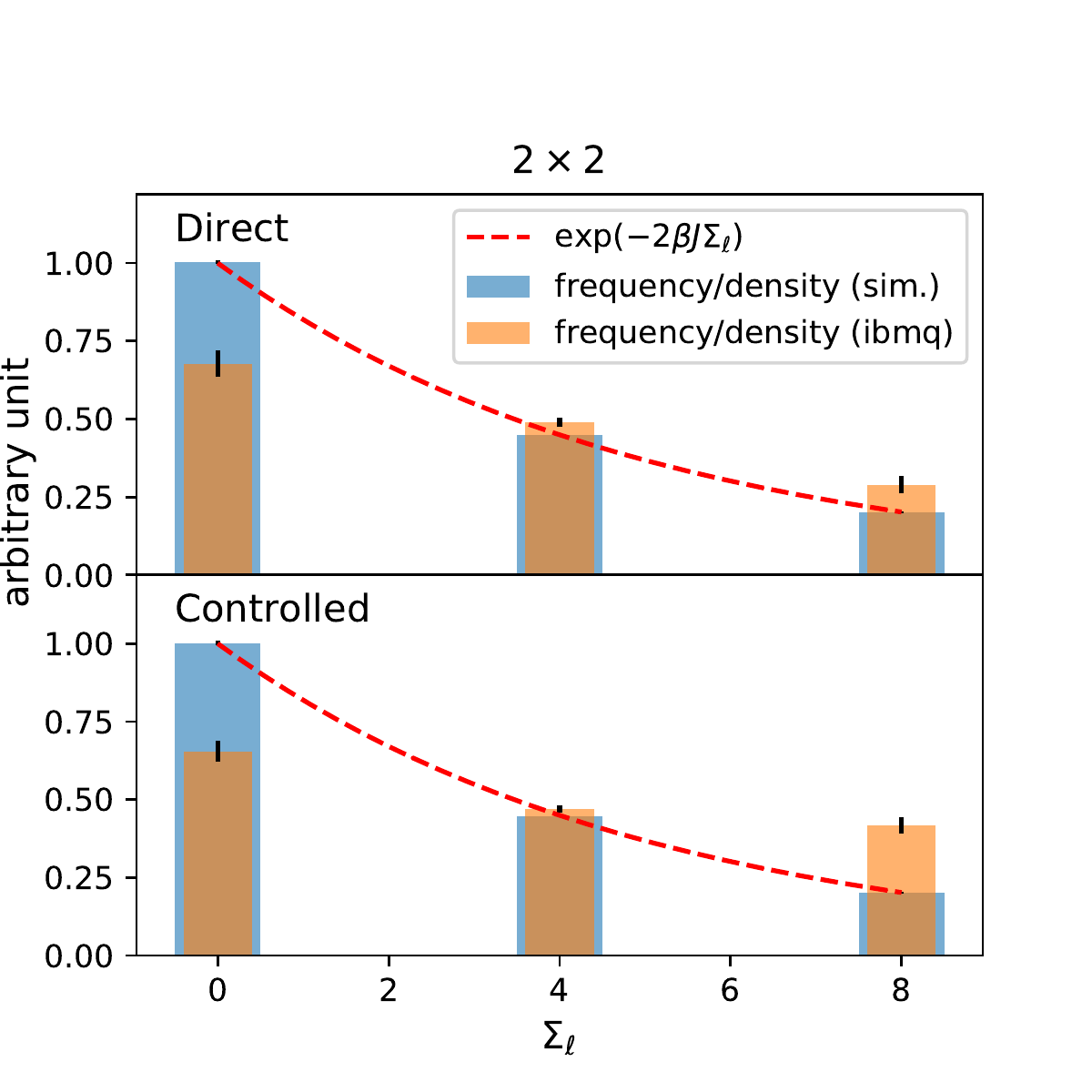}
    \caption{Distributions of $\Sl$ in the $2 \times 2$ Ising system for $\beta J=0.1$ obtained from quantum circuit simulation and the quantum hardware \texttt{ibmq\_cambridge}, running the direct (top) and controlled (bottom) amplitude transduction algorithms. Error bars represent statistical uncertainties.}
    \label{fig:boltzmann_2x2}
\end{figure}

\begin{figure}[t]
    \centering
    \includegraphics[width=0.48\textwidth]{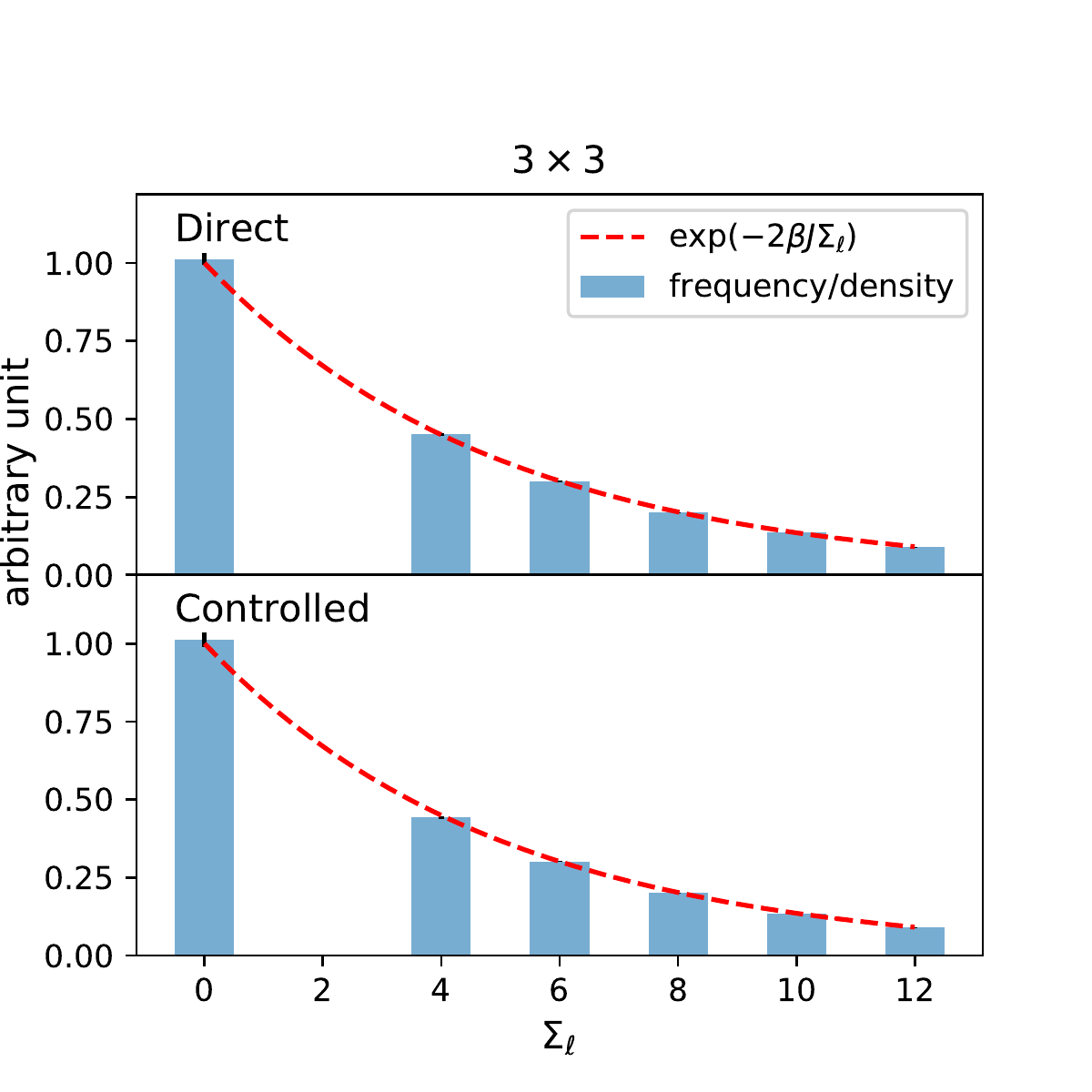}
    \caption{Distributions of $\Sl$ in the $3 \times 3$ Ising system for $\beta J=0.1$ obtained from quantum circuit simulation. See Fig.~\ref{fig:boltzmann_2x2} caption for details.}
    \label{fig:boltzmann_3x3}
\end{figure}

\begin{figure}[t]
    \centering
    \includegraphics[width=0.48\textwidth]{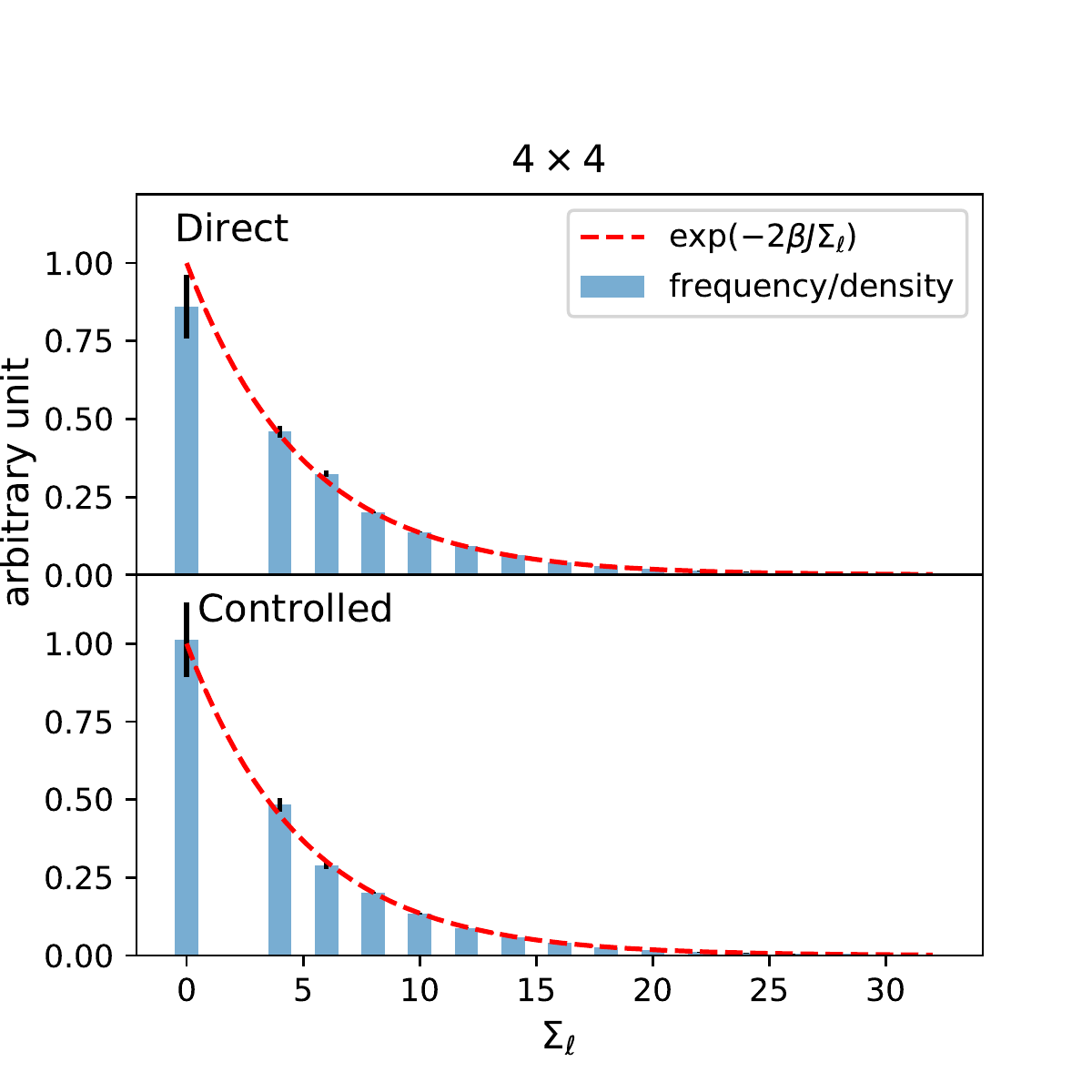}
    \caption{Distributions of $\Sl$ in the $4 \times 4$ Ising system for $\beta J=0.1$ obtained from quantum circuit simulation. See Fig.~\ref{fig:boltzmann_2x2} caption for details.}
    \label{fig:boltzmann_4x4}
\end{figure}

Figures~\ref{fig:boltzmann_2x2} to \ref{fig:boltzmann_4x4} show the resulting distributions of $\Sl$. Observed frequency of occurrence of each $\Sl$ is divided by the density of states (number of spin configurations resulting in each value of $\Sl$). The distributions from simulations follow the theoretical curve $\exp (-2\beta J \Sl)$, confirming that state preparation is successful. The distributions obtained from quantum hardware do not follow the theoretical curve, due to errors in CNOT, $X_{\pi/2}$, and measurement operations. The controlled variant has more of all three operations than the direct variant, and thus has a $\Sl$ distribution that is more dissimilar to the theoretical curve. It also has a greater discrepancy between $u^2$ and the efficiency in Table~\ref{tab:experiments}.

\begin{figure}[t]
    \centering
    \includegraphics[width=0.48\textwidth]{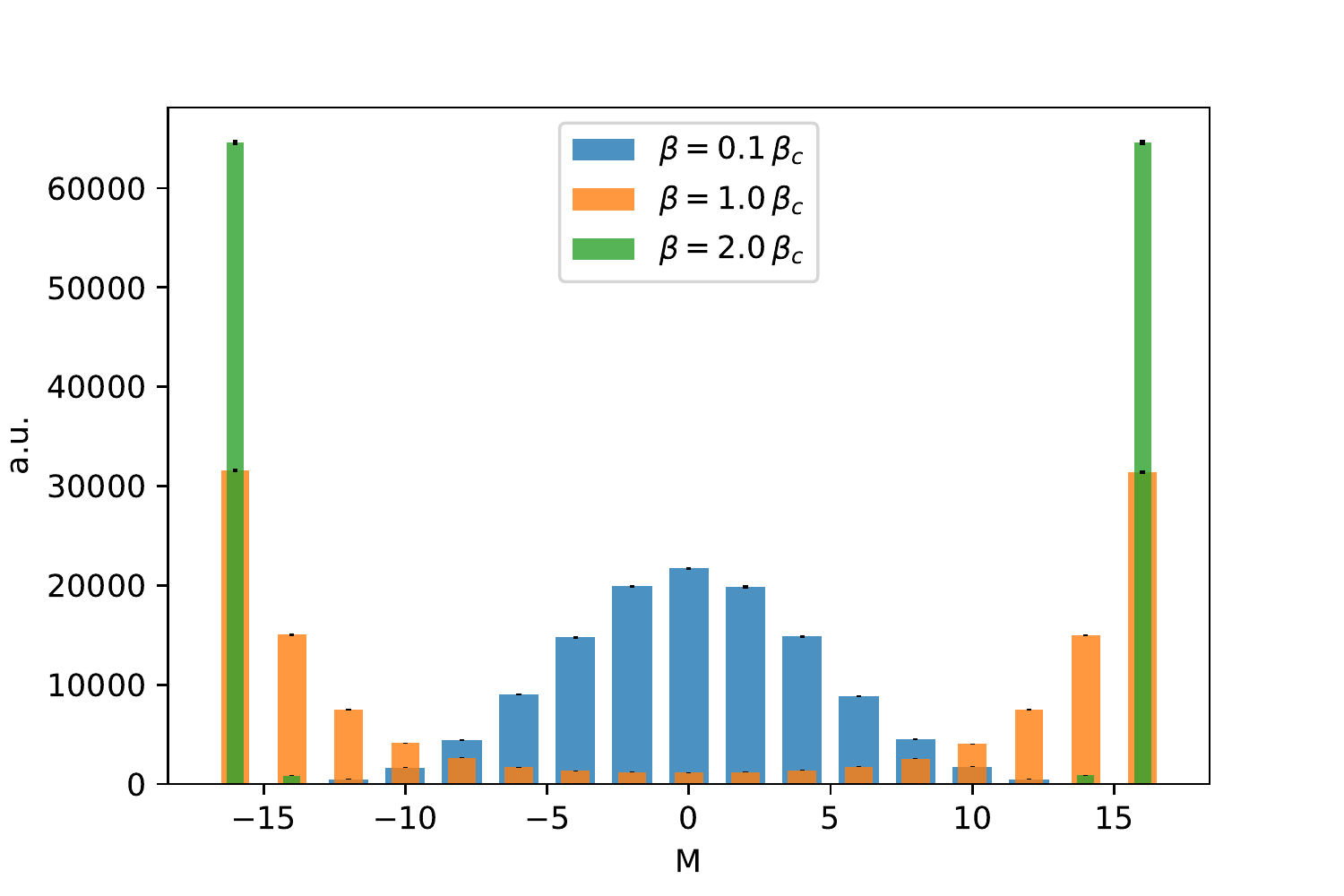}
    \caption{Distributions of magnetization $M$ of the $4\times 4$ Ising system under $\beta=0.1 \beta_c, \beta_c,$ and $2\beta_c$ obtained from quantum circuit simulation. Error bars represent statistical uncertainties.}
    \label{fig:magnetization}
\end{figure}

As an additional check, Fig.~\ref{fig:magnetization} shows the distributions of magnetization $M:=\sum_{i}s_i$ obtained from controlled-variant simulations of the $4 \times 4$ lattice with $\beta=0.1\beta_{c}, \beta_{c},$ and $2\beta_{c}$, where $\beta_{c} \sim 2.269/J$ is the inverse critical temperature of the square lattice Ising model at the thermodynamic limit ($N \to \infty$). At high temperature ($0.1\beta_{c}$), spins are more likely to be randomly oriented ($M \sim 0$), while at low temperature ($2\beta_{c}$), the entire lattice tends to be magnetized in one direction ($M \sim \pm N$). These results further confirm the validity of the probability distribution generated by the quantum circuit.

\section{Summary and outlook}
\label{sec:conclusions}

In this paper, we presented a novel method of quantum state preparation, where the amplitude of each computational basis state is transduced multiplicatively from the digitized logarithm of the target amplitude function, recorded in an auxiliary register. Two variants of the algorithm have been introduced. The direct variant is based on direct operations on the auxiliary register, thus requiring no additional qubits in amplitude transduction. Furthermore, the operations are implementable with single-qubit gates, making this variant fault-tolerant. The controlled variant employs another auxiliary register of the same number of qubits as the first one, but synthesizes the target state with a greater norm than the direct variant. This variant therefore requires less amplitude amplification operations than the other, potentially resulting in a smaller circuit depth overall.

Multiplicative amplitude transduction is demonstrated with a prototypical problem of generating spin configurations of a square lattice Ising model according to its Boltzmann distribution. The smallest-scale example circuits have been successfully executed on a real quantum hardware.

While the scale of the example problem run on the quantum hardware reported in this paper is limited mostly by the circuit depth, the dominant contribution to the complexity is from the spin-pair counting routine, i.e., the calculation of the logarithm of the amplitude. For problems with even simpler logarithm calculations, the simplicity of the algorithm is such that existing quantum computers with $\sim 50$ qubits may already be enough to put the algorithm in practical use. We anticipate that this will certainly be the case with near-future machines with $\mathcal{O}(100)$ qubits and lower error rates.

\paragraph{Code availability}

The Python code for the Ising model configuration generator is available as a Jupyter notebook at Ref. \cite{zenodo_2020_3866081}.

\bibliographystyle{unsrtnat}
\bibliography{main}

\end{document}